\documentclass[traditabstract]{aa}

\usepackage{natbib}
\usepackage{epsfig}
\usepackage{epsf}
\usepackage{color}
\definecolor{red}{rgb}{0.7,0,0}
\definecolor{blue}{rgb}{0,0,0.7}
\def\correc#1{{#1}}

\def\ergcms{erg~cm$^{-2}$~s$^{-1}$}
\def\nh{N$_\mathrm{H}$}
\def\cm2{cm$^{-2}$}

\def\integral{{\it{INTEGRAL}}}
\def\swift{{\it{Swift}}}

\def\chisq{$\chi^2_\nu$}
\def\deg{$^\circ$ }

\usepackage{longtable}
\usepackage[figuresright]{rotating}
\usepackage{lscape}
\usepackage{txfonts}
\begin{document}
   \title{\swift\ follow-up observations of 13 INTEGRAL sources}

   \author{J. Rodriguez
          \inst{1} 
          \and
          J.A. Tomsick \inst{2}
          \and 
          A. Bodaghee\inst{2}
}

   \offprints{J. Rodriguez}
\authorrunning{Rodriguez, et al. }
%\titlerunning{Swift }
   \institute{Laboratoire AIM, CEA/IRFU - CNRS/INSU - Universit\'e Paris Diderot, CEA DSM/IRFU/SAp,
 Centre de Saclay, F-91191 Gif-sur-Yvette, France\\
              \email{jrodriguez@cea.fr}
         \and
             Space Sciences Laboratory, 7 Gauss Way,
University of California, Berkeley, CA 94720-7450, USA\\
             }

   \date{}

% \abstract{}{}{}{}{} 
% 5 {} token are mandatory
 
  \abstract{The various IBIS/ISGRI catalogues contain a large population of hard X-ray sources whose nature 
is still unknown. Even if the $>20$ keV positional uncertainty provided by ISGRI is unprecedented, 
it is still too large to pinpoint the counterpart at other wavelengths, which is the only secure 
way of obtaining a source identification. We continue the work of trying to reveal the nature of 
these hard X-ray sources, starting with analysis of X-ray data collected via focusing X-ray telescopes, in
order to obtain arcsec accurate X-ray positions.  We can then identify counterparts at infrared and optical 
wavelengths and try to unveil the nature of the sources. We analysed data from observations of 13 \integral\ 
sources made with the \swift\ satellite. The X-ray images obtained by the X-Ray Telescope instrument 
allowed us to find possible counterparts to the IGR sources with a positional accuracy of a few arcsec. 
We then browsed the online catalogues (e.g., NED, SIMBAD, 2MASS, 2MASX, USNO B1.0) to search for counterparts 
at other wavelengths. We also made use of the X-ray spectral parameters in trying to identify the nature of those
objects. For the 13 objects, we found possible counterparts at X-ray energies and identified the IR/optical
and/or UV counterparts as seen with \swift/UVOT. We also discuss the likelihood of association of 
the X-ray and \integral\ source in each case. We confirm the previously proposed classification of IGR~J02524$-$0829 
(Sey 2 AGN),  J08023$-$6954 (RS CVn star), and J11457$-$1827 (Sey 1 AGN). For 7 of these sources
we give the first identification of their nature: IGR J02086$-$1742, J12060+3818, J12070+2535, 
J13042$-$1020, and J13412+3022 are AGN, and J14488$-$5942 is
a probable X-ray binary. For J03184$-$0014,  although we question the association of the IGR 
and \swift\ sources, we classify the latter as an AGN. We suggest that IGR J15283$-$4443 is 
a Galactic source, but we cannot  classify the source further. 
Finally, we question the association of IGR J11457$-$1827 and J23130+8608 with the X-ray sources we 
found, and go on to question the genuineness of the former IGR source. }

   \maketitle
%
%________________________________________________________________

\section{Introduction}
The most recent version of the IBIS catalogue contains more than 700 
hard X-ray sources \citep{bird10}. While a certain number were known as (hard)
X-ray emitters prior to the launch of \integral, about half of them have 
been detected for the first time above 20 keV with IBIS/ISGRI \citep{lebrun03}.
In this paper we  refer to these sources as `IGRs'\footnote{An up-to-date online 
catalogue of all IGRs can be found at http://irfu.cea.fr/Sap/IGR-Sources/ note the new 
address for the site}.  \citet{arash07} has collected known parameters (e.g., the 
absorption column density, \nh, the pulse period for Galactic sources with X-ray 
pulsations, the redshift for AGN, etc.) of all sources detected by \integral\ during 
the first four years of activity. With this they could study the parameter spaces
occupied by different families of sources and therefore  deduce 
important aspects of the physics of high-energy sources.  
However, many of these IGRs have still not been identified, and therefore 
any attempt to study, understand, and model populations of high-energy 
sources will be incomplete. The determination of the nature of these object 
is therefore extremely important if one wants to have the most complete 
view of the content of our Galaxy and our Universe. \\
\indent In this paper, we continue our work of identifying the unknown IGRs
that we started soon after the discovery of the first IGRs.  A first step
is to provide an $\sim$arcsec position with soft X-ray telescopes such as 
{\it {XMM-Newton}} \citep{rodrigue03_1632, rodrigue06_1632, bodaghee06}  {\it{Chandra}} 
\citep{john06_4igr, john08_igr,john09_igr}, and also 
\swift\ \citep{rodrigue08_igr,rodrigue09_igr, rodrigue09_19294}. We then search for 
counterparts at a position consistent with the refined X-ray position of 
a given source. As in \citet{rodrigue08_igr} and \citet{rodrigue09_igr} (Papers 1 and 2 
in the remainder of this article), we report here 
the analysis of \swift\ observations (XRT imaging and spectral analysis 
and UVOT imaging) of 13  IGRs that still lacked precise arcsec X-ray positions at the time 
of the writing of the paper. We 
also present the identification of IR and optical counterparts obtained from 
online catalogues such as SIMBAD, the United States Naval Observatory (USNO), 
the 2 Micron All Sky Survey point source and extended source 
catalogues\footnote{http://www.ipac.caltech.edu/2mass/} \citep[2MASS and 2MASX][]{skrutskie06}, 
and the NASA/IPAC Extragalactic Database 
(NED\footnote{http://nedwww.ipac.caltech.edu/index.html}). 
Although the presence of a bright \swift\ 
source within a given \integral\ error circle usually renders the 
association between the two sources likely, there is a slight probability that 
the two sources are not associated, especially in the case of dim X-ray sources. 
This is, also, exemplified by 
the few cases where several \swift\ sources are found within the \integral\ 
error circle.  Given the wide range of association probabilities from possible associations to 
nearly certain associations, no general statement can be given for the probability of associations. 
A low Galactic latitude source will have a higher chance of spurious association than a high latitude one.
For all sources, we discuss the likelihood of association
between the \integral, \swift, and counterparts at other wavelengths. 
Dubious cases (such as multiple possible counterparts) are discussed in more detail.  \\
\indent We start by introducing the \swift\ observations, and we briefly 
present the data reduction techniques in Sect.~2.  We then 
give the results of X-ray (Sect.~3) and IR/Optical/UV (Sect.~4) candidate counterparts identification. 
In Sect.~5 we describe the results for each source, including the results of the X-ray spectral analysis, 
 and discuss their possible nature.  We conclude the paper by summarising the results in Sect.~6.
\begin{table}
\caption{Journal of the \swift\ observations analysed in this paper.}
\begin{tabular}{lllll}
\hline
\hline
Source Id & Id  & Date Obs & Tstart & Exposure\\
(IGR)     &      &          &  (UTC) &  (s) \\
\hline
J02086$-$1742  &  00038021001  &  2009-09-10 & 14:08:56  &  4938 \\ 
J02524$-$0829  &  00036970001  &  2008-01-27 & 00:41:02  &  11117  \\ 
                     &  00036970002  &  2008-06-11 & 00:51:48  &  4310  \\ 
J03184$-$0014  &  00030995001  &  2007-11-07 & 00:12:58  &  9192 \\ 
                        &   00036969001 & 2008-02-28  & 01:48:52 & 8378 \\
                        &   00036969002 & 2008-02-29  & 19:32:52 & 4711 \\
                       &   00036969003 & 2009-06-24  & 19:25:35 & 1030 \\
J08023$-$6954  &  00036095001  &  2006-12-27 & 07:45:23  &  701  \\ 
                       &  00036095002  &  2006-12-30 & 08:04:22  &  5736  \\ 
J11457$-$1827 & 00035645001 & 2006-07-27  &  00:05:11 & 10000 \\
                        & 00035645002 & 2006-07-29 &  00:19:43  & 3384 \\
J12060+3818    & 00037838001 & 2008-10-13 &  00:31:19 & 5363 \\
J12070+2535    & 00037837001 & 2008-10-27 & 18:03:14 & 2032 \\
J12482-5828  &  00038349001  &  2008-12-19 & 00:43:32  &  3718 \\ 
                     &  00038349002  &  2009-05-08 & 01:06:36  &  1847  \\ 
J13042-1020  &  00031153001  &  2008-03-03 & 15:34:26  &  2948  \\ 
                     &  00031153002  &  2008-03-05 & 14:11:09  &  816  \\ 
                     &  00031153003  &  2008-03-05 & 14:14:18  &  4889 \\ 
                     &  00031153004  &  2008-03-07 & 00:21:55  &  185  \\ 
                     &  00031153005  &  2008-03-07 & 00:22:18  &  6621  \\ 
                     &  00031153006  &  2008-03-09 & 11:45:39  &  2676  \\ 
                     &  00031153007  &  2008-03-13 & 13:44:13  &  3327  \\ 
                     &  00031153008  &  2008-03-16 & 11:00:24  &  3279  \\ 
                     &  00031153009  &  2008-04-23 & 05:01:41  &  3246  \\ 
J13412+3022 &  00037380001  &  2008-08-24 & 00:42:39  &  2694  \\ 
                     &  00037835001  &  2008-08-25 & 15:17:06  &  3562  \\ 
J14488-5942  &  00039094001  &  2009-09-25 & 19:25:22  &  16413  \\ 
J15283-4443  &  00036114001  &  2007-01-06 & 07:36:11  &  5534  \\ 
J23130+8608  &  00037078002  &  2007-07-09 & 01:33:08  &  7908  \\ 
                      &  00037078003  &  2007-07-11 & 00:03:00  &  11191  \\ 

\hline
\hline
\end{tabular}
\label{tab:log}
\end{table}

\section{Observations and data reduction}
We searched the \swift\ archive for observations at less than 
10\arcmin\ of any IGR that has a position uncertainty 
greater than about 10\arcsec. We excluded most of the new IGRs
found in the Galactic centre by \citet{bird10}, as too many possible 
X-ray counterparts can be found in the IBIS error. We also excluded sources 
for which a clear positional uncertainty is not given in \citet{bird10}, because this 
may indicate source confusion in IBIS. 
We then only retained the pointings during which the XRT instrument 
was in photon-counting mode since it is the only mode that provides a fine position.  
We report on the results of the 13 sources for which we found an X-ray source 
within the IBIS error box.  The observing log for these is reported in Table~\ref{tab:log}. \\
\indent We reduced the \swift\ data with the {\tt{HEASoft V6.7}} 
software package and the calibration files issued on 2009 December 1 and 2009 October
7 for the UVOT and XRT instruments, respectively.  The reduction procedure is  
identical to those presented in Papers 1 and 2.  \correc{For XRT, level 2 cleaned event files were 
obtained with {\tt{xrtpipeline}} with standard parameters\footnote{see http://heasarc.gsfc.nasa.gov/docs/swift/analysis/}. } \\
\indent The XRT individual pointings of a given source were \correc{then} co-added  
with {\tt{xselect}}. We extracted
spectra and light curves with {\tt{xselect}} from a circular region with a radius 
of 20 pixels centred on the best position, while we obtained the background 
products from a source-free circular region with a radius of 40 pixels.  
The presence of columns of dead pixels in the XRT meant we 
had to produce ``true'' exposure maps \correc{with {\tt{xrtexpomap}} that were given as input 
to {\tt{xrtmkarf}} }to produce corrected ancillary response 
files. We rebinned the spectra to have at least 20 counts 
per channel, which allows for $\chi^{2}$-minimisation in the fitting with
{\tt{XSPEC 12.5.1}}. When this criterion was not achievable, the Cash 
statistic \citep[hereafter C-statistic][]{cash76} was used instead.\\
\indent When available, we analysed the UVOT level 2 data obtained from 
the \swift\ data archive.  We first corrected the aspect for each individual 
UVOT exposure with the {\tt{uvotskycorr}} tool, calculating the aspect correction 
via comparison to the USNO-B1.0
catalogue\footnote{http://tdc-www.harvard.edu/software/catalogs/ub1.html}\citep{monet03}.
Then, we summed the aspect-corrected individual exposures \correc{and individual exposure maps} 
with {\tt{uvotimsum}} and performed the UVOT astrometry \correc{using the summed images and exposure
maps} with the {\tt{uvotdetect}} tool.  \correc{The tool {\tt{uvotsource}} was finally used to obtain the photometry 
of the sources.  The UVOT magnitudes were estimated on a source region of 5\arcsec\ radius centred 
on the best source position obtained with  {\tt{uvotdetect}}, using a region of 20\arcsec\ radius free of sources as 
background. This tool corrects the source count rates for coincidence losses, makes an aperture correction, 
and converts the results to magnitudes following $m_{\rm{source}}=Z_{pt}-2.5\times log(CR)$ \citep{poole08}, where 
$Z_{pt}$ is the photometric zero-point of the filter considered, and $CR$ is the corrected source count rate. 
All corrections have been applied with the UVOT calibration files released on 2009 October 7. 
The complete description of the UVOT photometric calibration can be found in \citet{poole08}.}

\section{X-ray astrometry with \swift/XRT}
For each source we produced an image accumulating the maximum number 
of pc mode pointings available. We then searched for potential X-ray counterparts 
within the IBIS error box with {\tt {XIMAGE}}. We retained only sources that had a 
signal-to-noise ratio (SNR) greater than 4$\sigma$ and went down to 3$\sigma$ when 
no significant excesses were found. We estimated the best source position and 
errors with {\tt{xrtcentroid}}.  The list of candidates counterpart, their number within 
the IBIS error box, X-ray position, and 
SNR are reported in Table~\ref{tab:xray}.

\begin{table*}
\caption{List of sources for which X-ray counterparts have been found}
\begin{tabular}{llcccccccc}
\hline
Name (IGR J) & \#  & RA & Dec & error & l & b& det. sig & $90\%$  IBIS   & Offset from the\\
                      &      &      &        & (\arcsec) &  &   &    ($\sigma$)        &  confidence radius & IBIS position\\
                      &      &      &        &               &  &   &            &                   (\arcmin)       & (\arcmin) \\  
\hline
J02086$-$1742  & 1 & 02h 08m 34.8s & $-$17\deg\ 39\arcmin\ 33.3\arcsec & 3.3 & 188.9293\deg\ & $-$69.8490\deg\ & 28  & 4.5 & 3.6 \\ 
J02524$-$0829  & 1 & 02h 52m 23.4s & $-$08\deg\ 30\arcmin\ 38.5\arcsec & 3.6 & 185.5565\deg\ & $-$55.8849\deg\ & 22 & 2--3 & 1.9 \\
J03184$-$0014  & 1 & 03h 18m 17.4s & $-$00\deg\ 17\arcmin\ 47.9\arcsec & 5.2 & 181.8103\deg\ & $-$45.7089\deg\ & 4.3 & 4 & 4.1\\
J08023$-$6954  & 1 & 08h 02m 41.8s & $-$69\deg\ 53\arcmin\ 38.7\arcsec & 6.3 & 282.6101\deg\ & $-$19.5480\deg\ & 3.2 & 2--6 & 2.3\\
J11427+0854 & 1 & 11h 42m 27.1s & +08\deg\ 54\arcmin\ 48.0\arcsec & 5.6 & 257.8755\deg\ & 65.5549\deg\  & 3.4 & 3.9 & 3.2\\
J11457$-$1827$^\dagger$ & 1 & 11h 45m 40.5s & $-$18\deg\ 27\arcmin\ 15.5\arcsec & 3.5 & 281.8546\deg\ & 41.7102\deg\ & 100 & 3.4 & 0.3\\
J12060+3818 & 1 & 12h 06m 17.2s & +38\deg\ 12\arcmin\ 37.0\arcsec & 4.75 & 160.6109\deg\ & 75.4265\deg\ & 6.6 & 4.0$^\ddagger$ & 6.0\\
J12070+2535 & 1 & 12h 07m 05.3s & +25\deg\ 39\arcmin\ 06.1\arcsec & 4.5 & 218.8913\deg\ & 79.9624\deg\ & 8.3 & 3.4$^\ddagger$ & 4.2 \\
J13042$-$1020  & 4 & 13h 04m 14.2s & $-$10\deg\ 20\arcmin\ 20.7\arcsec & 3.7 &308.0957\deg & 52.4043\deg & 15 & 4.1 & 1.0 \\
\hspace*{0.55cm} " &   & 13h 04m 13.6s & $-$10\deg\ 21\arcmin\ 21.9\arcsec & 3.8 & 308.0894\deg & 52.3875\deg & 6.2   & " & 1.3\\
\hspace*{0.55cm} " &   & 13h 04m 11.6s & $-$10\deg\ 19\arcmin\ 52.3\arcsec & 3.7 & 308.0793\deg & 52.4130\deg & 5.2  &" & 1.7\\
\hspace*{0.55cm} " &   & 13h 04m 09.8s & $-$10\deg\ 19\arcmin\ 44.4\arcsec & 4.0 & 308.0676\deg & 52.4158\deg & 4.9  &" & 2.2\\
J13412+3022$^\star$  & 1 & 13h 41m 11.2s & +30\deg\ 22\arcmin\ 40.9\arcsec & 4.25 & 52.4662\deg & 78.6298\deg & 8.8 & 3.8 & 2.1\\
J14488$-$5942  & 2 & 14h 48m 43.3s & $-$59\deg\ 42\arcmin\ 16.3\arcsec & 3.7 & 317.2340\deg & $-$0.1298\deg & 16 & 4.4 & 0.7\\
\hspace*{0.55cm} " &   & 14h 49m 00.5s & $-$59\deg\ 45\arcmin\ 03.9\arcsec & 4.9 & 317.2462\deg & $-$0.1875\deg & 3.8 &" & 3.3\\
J15283$-$4443  & 1 & 15h 28m 16.1s & $-$44\deg\ 43\arcmin\ 41.7\arcsec & 6.0 & 330.0392\deg & 9.7322\deg & 3.3 & 3 & 1.4\\
J23130+8608 & 1 & 23h 08m 55.3s & +86\deg\ 05\arcmin\ 50.9\arcsec & 4.8 & 121.0910\deg & 23.5978\deg & 3.8 & 4.8 & 4.7\\
\hline
\end{tabular}
\begin{list}{}{}
\item[$^\dagger$]From \citet{winter09}.
\correc{\item[$^\ddagger$]IBIS error possibly underestimated (see text).}
\correc{\item[$^\star$]Wrong name (J13415+3033) given in \citet{bird10}.}
\end{list}
\label{tab:xray}
\end{table*}

\section{Counterparts at other wavelengths}
We searched the NED, 2MASS, 2MASX,  
and the USNO-B1.0 online catalogues and UVOT images for infrared, 
optical, and UV counterparts within the \swift/XRT error circle of each of the potential
X-ray counterpart reported in Table~\ref{tab:xray}.  
Infrared counterparts that are newly identified from this search are reported 
in Table~\ref{tab:ircounterparts}. The typical positional accuracy for 
the 2MASS sources is 0.5\arcsec\ \citep{skrutskie06}, while that of the USNO-B1.0 
sources is typically 0.2\arcsec\ \citep{monet03}. The magnitudes and UV 
positions of the optical and UV counterparts are reported in 
Table~\ref{tab:uvcounterparts}. The USNO-B1.0 photometric 
accuracy is typically 0.3 mag \citep{monet03}. The lower limits on the UVOT 
magnitudes are given at the $3\sigma$ level. The UVOT positional uncertainties 
are dominated by a 0.5\arcsec\ systematic uncertainty (90\% confidence) for 
each source. 

\begin{table*}[htbp]
\caption{List of newly identified infrared counterparts in the 2MASS and 2MASX catalogues.}\label{tab:ircounterparts}
\begin{tabular}{llcccc} 
\hline\hline             
Name & Counterpart &\multicolumn{3}{c}{Magnitudes} & Offset from the\\
(IGR)  &  & J & H & K$_\mathrm{s}$  & XRT position (\arcsec)\\ 
J02086$-$1742       & 2MASS J02083494$-$1739347 & $14.51\pm0.04$& $13.61\pm0.04$ & $12.54\pm0.03$ & 2.6\\       
J02524$-$0829       & 2MASX J02522337$-$080376   & $11.79\pm0.02$ & $10.99\pm0.02$ & $10.66\pm0.04$ & 0.9\\
J03184$-$0014$^\star$       & 2MASS J03181753$-$0017502 & -- -- -- & -- -- -- & $15.22\pm0.15$ & 3.1\\
J08023$-$6954       & 2MASS J08024164$-$6953377 & $12.25\pm0.02$ & $11.65\pm0.02$ & $11.52\pm0.02$& 1.3\\
J11457$-$1827$^\dagger$ & 2MASX J11454045$-$1827149 & $12.87\pm0.03$ & $12.15\pm0.04$ & $11.65\pm0.06$ & 0.9\\
J12070+2535         & 2MASS J12070528+2539059    & $15.53\pm0.06$ & $14.71\pm0.05$ & -- -- -- & 0.3\\ 
J13042$-$1020 \#1 & 2MASX J13041438$-$1020225 & $9.62\pm0.02$ & $8.50\pm0.02$ & $8.68\pm0.02$ & 3.4\\
J13412$-$3022       & 2MASX J13411117+3022411   & $12.03\pm0.02$ & $11.33\pm0.03$ & $11.02\pm0.04$ & 0.4\\
J14488$-$5942\#1  & 2MASS J14484322$-$5942137   & $15.46\pm0.07$ & $13.53\pm0.04$ & $12.43\pm0.04$& 2.6\\
J15283$-$4443       & 2MASS J15281596$-$4443416   & $10.34\pm0.02$ & $10.03\pm0.02$ & $9.97\pm0.02$ & 1.5\\
J23130+8608          & 2MASS J23085980+8605526 & $12.98\pm0.02$& $12.40\pm0.03$  & $12.22\pm0.02$ & 4.9\\
\hline

\hline
\hline
\end{tabular}
\begin{list}{}{}
\item[$^\star$]See also Paper 2.
\item[$^\dagger$]From \citet{winter09}.
\end{list}
\end{table*}

\section{X-ray spectral analysis and possible identifications of the thirteen IGRs}
In this section we report the results for each source and discuss 
their possible nature. We also include the results of the X-ray spectral analysis 
of the XRT data. In all cases, we started to fit the source spectra 
with a simple model of an absorbed power-law. This provided an acceptable 
representation in the large majority of the cases. The absorption due to 
intervening material along the line of sight is  obtained from the 
Leiden/Argentine/Bonn (LAB)  surveys of Galactic H~I in the 
Galaxy\footnote{http://www.astro.uni-bonn.de/$\sim$webaiub/english/tools\_labsearch.php}. 
The LAB Survey is the most sensitive Milky Way H~I survey to date, with the most 
extensive coverage both spatially and kinematically and an angular resolution of 
0.6 degrees \citep{kalberla05}. \\
\indent The spectral parameters we 
obtained are reported in Table~\ref{tab:spectral}.  The errors on the X-ray spectral 
parameters (including upper limits) are given at the 90$\%$ confidence level.
 \correc{We also report there  
the extrapolated hard X-ray fluxes of these sources as measured by IBIS and obtained by extrapolating 
the best XRT spectrum in the same energy range. Note that while, in most cases, these fluxes 
are estimated over the 20--40 keV range, in some cases, they are obtained in different ranges to match
published results on those sources. }
 To estimate the luminosity of the  candidate AGN we used H$_0$=75~km/s/Mpc 
to convert the redshift (of the suggested counterpart) to distance.

\subsection{\object{J02086$-$1742}}
This source is  one of the new IGRs reported in the most recent version of the 
IBIS catalogue \citep{bird10}, for which inspection of the \swift\ archives shows  an observation 
aimed at a \swift\ object.  Although the coordinates 
of \object{Swift J0208.5$-$1738} and that of the IGR sources indicate an offset of about 
5\arcmin, both objects have error boxes that render their association likely. There is 
one single bright X-ray source with a position  that is compatible with both
the \swift\ and \integral\ ones indicating that the sources are the same  (Table~\ref{tab:xray}). 
The X-ray position is also at 22\arcsec\ from the position of  \object{1RXS J020835.8$-$173950}. Given
the 20\arcsec\ error box given by {\em{Rosat}}, it is likely that the XRT source and the {\em{Rosat}} one are the same. 
The XRT error box contains \object{NVSS J020834$-$173933} a $\sim 42$ mJy 1.4 GHz radio source 
as reported in NED.  The source is also clearly detected in the infrared, optical and UV bands 
(Tables~\ref{tab:ircounterparts} and \ref{tab:uvcounterparts}).\\
\indent The XRT spectrum is represented well by a hard power-law model
with very little absorption (\chisq=1.05 for 41 dof) compatible with the value 
on the line of sight obtained from the LAB survey \citep{kalberla05}. The extrapolated 20--40 keV flux 
is compatible with the flux measured by IBIS \citep{bird10}, which further strengthens 
the associations of the X-ray source and the IBIS one. The high galactic latitude, the radio 
detection, and the brightness in all IR/optical and UV bands as well as the detection at soft and 
hard X-ray energies argue in favour of the source being an AGN. The low value of the absorption 
would argue for of a type 1 AGN. \correc{\citet{masetti10} also classify the source as a 
Sey 1.2 AGN through optical spectroscopy.}

\begin{landscape}
%\centering
\begin{table}

\caption{Magnitudes and UVOT positions of the newly identified optical 
and UV counterparts in the USNO-B1.0 catalogue (I, R, and B bands) and \swift/UVOT detector (V, B, U, UVW1, UVM2, and UVW2  bands). }
\begin{tabular}{llccclllllllll}
\hline
\hline
Name & USNO B1.0 & \multicolumn{2}{c}{UVOT position$^{\dagger,a}$}  & \multicolumn{8}{c}{Magnitudes$^{a,b}$} \\
(IGR) &        & RA      & DEC                  &  I & R & V & B$^c$ & U & UVW1 & UVM2 & UVW2 \\
\hline
J02086$-$1742      & 0723-0034272 & 02h 08m 34.9s & $-$17\deg\ 39\arcmin\ 34.8\arcsec & 14.2 & 15.0 & -- -- -- & 16.4 & $14.53\pm0.03$& -- -- -- & -- -- --& -- -- --\\ 
J02524$-$0829$^\P$      & 0814-0027101$^\star$ & 02h 52m 23.4s & $-$08\deg\ 30\arcmin\ 37.8\arcsec & 10.7 & 11.3 & -- -- -- & 11.4 & $16.88\pm0.03$ &-- -- -- & $18.93\pm0.05$ & -- -- -- \\
J08023$-$6954      & 0201-0188880             & 08h 02m 41.6s & $-$69\deg\ 53\arcmin\ 38.0\arcsec & 12.9 & 14.3 & -- -- -- & 15.6 & -- -- -- & $18.29\pm0.08$$^\ddagger$ & $20.75\pm0.37$ & $19.3\pm0.15$$^\ddagger$ \\
J11427+0854& No counterpart   & 11h 42m 27.2s & 8\deg\ 54\arcmin\ 47.9\arcsec  & -- -- -- & -- -- -- & -- -- -- & -- -- -- & -- -- -- & $18.86\pm0.07$$^\ddagger$ & $19.06\pm0.06$ & -- -- -- \\
J11457$-$1827     & 0715-0241351$^\star$  &-- --  -- &-- --  -- & 11.2 & 10.5 & -- -- -- & 10.6 & -- -- -- & -- -- -- & -- -- --& -- -- -- \\
J12060+3818        & 1282-0243241 & 12h 06m 17.4s & 38\deg\ 12\arcmin\ 35.3\arcsec & 18.4 & 18.9 & $18.7\pm0.2$ & $19.0\pm0.1$ & $18.47\pm0.09$ & $18.07\pm0.07$ & $17.97\pm0.07$ & $18.47\pm0.06$\\
J12070+2535$^\P$        & 1156-0188536$^\star$ & 12h 07m 05.3s & 25\deg\ 39\arcmin\ 06.5\arcsec & 16.4 & 16.1 & $17.5\pm0.1$ & $18.2\pm0.1$ & $17.9\pm0.1$ & $18.6\pm0.1$ & $18.9\pm0.2$ & $19.1\pm0.1$\\ 
J13042$-$1020 \#1$^{\ddagger,\P}$  & 0796-0240939 & 13h 04m 14.3s & $-$10\deg\ 20\arcmin\ 21.1\arcsec & 6.7 & 11.2 & $14.35\pm0.01$  & $15.33\pm0.01$ &  $15.66\pm0.02$ & $16.80\pm0.02$ & $17.55\pm0.03$ & $17.41\pm0.02$\\
J13412+3022$^\P$        & 1203-0214236$^\star$  & 13h 41m 11.2s & 30\deg\ 22\arcmin\ 41.5\arcsec & 9.2 & 9.7 & $14.96\pm0.03$ & $15.91\pm0.03$ & $16.15\pm0.04$ & $16.93\pm0.07$ & $17.29\pm0.05$ & $17.38\pm0.05$\\
J15283$-$4443     &7847-00975-1& 15h 28m 16.0s & $-$44\deg\ 43\arcmin\ 42.0\arcsec & 9.9 & 10.3 & \multicolumn{3}{c}{\correc{Detector saturated}} & $12.58\pm0.04$ & $13.06\pm0.04$ & $13.25\pm0.04$\\
J23130+8608        & 1760-0040693 & 23h 08m 59.8s & 86\deg\ 05\arcmin\ 53.0\arcsec & 12.8 & 14.4 & -- -- -- & 16.3 & $17.51\pm0.03$ & $18.69\pm0.16$ & $21.15\pm0.21$ & -- -- -- \\
\hline
\label{tab:uvcounterparts}
\end{tabular}
\begin{list}{}{}
\item[$^\dagger$]UVOT positional accuracy dominated by a statistical uncertainty of 1.1\arcsec.
\correc{\item[$^\P$]UVOT source (possibly) extended; UVOT magnitudes not calculated over the  extension of the source.} 
\item[$^\ddagger$]UVOT magnitudes averaged over multiple pointings.
\item[$^\star$]Multiple USNO-B1.0 sources in XRT error circle. This is the closest to the IR source.
\item[$^a$] The long dashes indicate the absence of corresponding data.
\correc{\item[$^b$]Magnitudes with no errors are obtained from the USNO B1.0 catalogue and have a typical uncertainty of 0.3 mag.}
\item[$^c$] Obtained from the UVOT detectors when available.
\end{list}

\end{table}
\end{landscape}
\subsection{\object{J02524$-$0829}}
This object was first reported by \citep{krivonos07} and associated to MCG-02-08-014, a z=0.016721
Sey 2 galaxy \citep{bikmaev08}; however, no refined X-ray position was published for this 
object, leaving the possibility of a chance association. We found one single X-ray source within the 
IBIS 3\arcmin\ error box. The XRT position (Table~\ref{tab:xray} is at 1\arcsec\ from the position of 
the Sey 2 galaxy, which confirms the association of these sources.  The X-ray position is also clearly compatible with the nucleus of a spiral galaxy seen in the UVOT images 
(not shown). The magnitudes measured in the U-filter show some slight variability between the two pointings (it has 
$m_U=15.373\pm0.08$ during the second one). \\
\indent The X-ray spectrum is well-fitted with an absorbed power-law (\chisq=1.3 dof 27 dof). 
The absorption significantly exceeds to the value measured on the line of sight, which indicates
that the source is significantly absorbed, while the photon index, although quite poorly constrained, 
 has a value typical of a type 2 AGN.
All our results confirm that IGR J02524$-$0829 is associated with MCG-02-08-014, making it  
a  Sey 2 AGN.

\subsection{\object{J03184$-$0014}}
This very poorly \correc {studied} object was first reported in \citet{bird07}, and a single \swift\ observation
was analysed by us in Paper 2. We  found a single X-ray source, however, outside the IBIS 
error, which led us to dismiss it as the counterpart to the IGR source (Paper 2). Here we used 
3 additional \swift\ observations, which allowed us to obtain a total exposure on this region that is 
about 2.5 times longer than in Paper 2. \\
\indent As previously reported, there is no X-ray source within the 4\arcmin\ IBIS error box 
of this source. The most significant X-ray source in this field is 4.1\arcmin\ from the best 
position and is consistent with the object mentioned in Paper 2, and the position we report 
here should be considered as the most accurate one. The IR counterpart, 
\object{2MASS J03181753$-$0017502} (see Paper 2) is  classified as a  quasar in SIMBAD. We, however, 
point out that this classification is based on the positional coincidence of a Quasar (dubbed SDSS J03184$-$0015)
within the IBIS error box mentioned in \citet{bird07}. This object is not associated to \swift\ and to the 2MASS source 
we report here either. Instead, the X-ray error box 
contains \object{SDSS J031817.53$-$001750.0} reported as an extended source and classified as 
a galaxy in NED.  There is no source in the USNO B1.0 catalogue within the XRT error box. There is no obvious 
UVOT counterpart in the two pointings during which the X-ray position in the UVOT field of view (fov). The source 
position is, however, close to a very bright source and confusion with this nearby object cannot 
be completely excluded. \\
\indent The X-ray spectrum has a very low statistical quality. It is fitted with an absorbed
power-law (C=23 for 27 dof), but the parameters are poorly constrained (Table~\ref{tab:spectral}).
The parameters are compatible with any X-ray emitting source, and the extrapolated 20--40 keV 
flux is not compatible with the 20--40 keV flux from \citet{bird07}. \\
\indent The X-ray position of the source may indicate that it is an active galaxy, although the low statistical 
quality of the data leads to some caution. In Paper 2 we did not favour an association of the \swift\ and
IGR source. The finding of a Galaxy within the \swift\ error box may, however, question this conclusion. 
In any case we tentatively classify the \swift\ source as 
an active galaxy and suggest that it could be associated to the IGR. In this case, the source is not 
the quasar mentioned in SIMBAD. One should also keep in mind that the IGR object could be a spurious 
IBIS source since it is not reported in the last IBIS catalogue of \citet{bird10} and 
\citet{krivonos07}\footnote{see also http://hea.iki.rssi.ru/rsdc/catalog/index.php 
for an updated version of the \citet{krivonos07} catalogue}.  

\subsection{\object{J08023$-$6954}}
This source was discovered by \citet{revnivtsev06}, \correc{who have already suggested the source was an
 chromospherically active star (RS CVn type). Optical spectroscopy led  \citet{masetti08} to confirm 
the suspected type of this object.} These authors, however,  point out that this identification 
is only tentative since they had just a marginal, possible X-ray counterpart. Re-analysis of the 
\swift\ data shows the presence of a $3.2\sigma$ source within the error box. The X-ray position is compatible 
with that of the counterpart analysed by \citet{masetti08}. The XRT error box contains 
IR, optical, and UV counterparts with positions that are compatible. This gives strength to the reality 
of the X-ray source.  The UVW1 magnitudes had compatible values 
between the two pointings. The average of these are reported in Table \ref{tab:uvcounterparts}.\\
\indent  The spectrum has an extremely low statistical quality (only 13 net counts), and basically has  
no count above 2 keV, which prevents any sound spectral analysis. That all source counts are below 2 keV 
indicates that this source is not highly absorbed. This and the detection of the source in the UV and
 optical domains may indicate a nearby source.  \correc{A tentative fit of the 0.2--2 keV spectrum shows that 
a simple power-law with $\Gamma=2.4_{-0.7}^{+0.8}$ provides a good representation (C=7.2 for 13 dof). The 2--10 keV
flux is $2.6\times10^{-14}$~\ergcms. The extrapolated 20--40 keV flux is at least a factor of 3400 lower than the flux reported by
\citet{revnivtsev06}.} 
This source is mentioned in the online version of the Krivonos' catalogue, but not in \citet{bird07} and \citet{bird10}. 
This may indicate a variable source that is compatible with the proposed RS CVn classification,
since these sources are known to be variable. We also note that  RS CVn are known X-ray emitters 
\citep[e.g.][]{osten07}, 
and they can sometimes be detected at hard X-rays, as exemplified by the \swift/BAT detection of the RS CVn star II Peg
\citep{barthelemy05,osten07} during a flare.  \correc{The detection of the latter is a short-lived event and any comparison
with the \integral\ source may be questioned. \citet{osten07} report their analysis of the full flare of II Peg as seen with \swift.
Hard X-ray detection of II Peg by the BAT telescope is visible over more than $8$~ks, and less than $12$~ks. Over this period, the 
10-200 keV flux is found around $1.2\times10^{-9}$~\ergcms\ \citep{osten07}, which is easily detectable by \integral\ in a short 
exposure. While such a flare would be missed in a total 2~Ms observation such as the one reported by \citet{revnivtsev06}, one should note 
that IGR J08023$-$6954 lies close to a border of the covered field. Using Fig. 1 of \citet{revnivtsev06} and  assuming they reached a  
0.16 mCrab sensitivity for the full 2 Ms exposure, one can roughly estimate an effective 
exposure of $\sim140$~ks at the position of the IGR source. A II Peg-like flare (assuming a constant flux during 10 ks) would be detected
by \integral\ during this accumulation time.  It is therefore possible that \citet{revnivtsev06}
caught the source during an outburst. Not-detecting this source in the long term IBIS catalogues is also 
compatible with the source being highly  variable and showing short-lived hard X-ray activity. }
\correc{We do not offer any definite proof, the source could be a fake detection by the IBIS 
telescope; if true, we suggest that the IGR source and the \swift\ one 
could possibly be associated, which would then confirm that the hard X-ray source is an RS CVn. }
 
\subsection{\object{J11427+0854}}
This source was first reported in \citet{paltani08}, who, however, give a low probability of 
genuineness.  They tentatively identify it with a 2MASX source. We found one single 
faint X-ray source within the IBIS error box even when extending 
the  error box to the $\sim5$\arcmin\ value expected for a $5\sigma$ source \citep{gros03}. There are no
2MASS or USNO B1.0 source within the \swift/XRT error box of this potential 
counterpart, and it lies at more than 4\arcmin\ from the 2MASX source 
mentioned in \citet{paltani08}. Therefore an association of the two is ruled 
out. The XRT error box contains \object{SDSS J114227.17+085448.2}, which is 
classified as a point source in NED. We find a UV counterpart in the UVW1 and UVM2 UVOT 
filters, which may indicate that this object emits mostly at short wavelengths.\\
\indent The X-ray spectrum has only 15 counts and is  consistent with no 
counts above 2 keV, which could indicate a very soft source. Given the 
low statistical quality, a fit cannot be performed. The low probability 
of reality for the \integral\ source, and the faintness of the \swift\ 
 renders the question of the association dubious,  although we cannot  
conclude much about the nature or reality of this source.

\subsection{\object{J11457$-$1827}}
This source was first reported in \citet{paltani08} who suggest an association with \object{1H 1142$-$178} based on
the location of the latter within the IBIS error box; therefore, the \integral\ source is a Sey 1 AGN. A \swift\ refined 
position of the source is given by \citet{winter09}, who associate the source to 
\object{2MASX J11454045$-$1827149} reported as a z=$0.0329\pm0.0001$ galaxy in NED.  We also 
found a possible counterpart in the USNO B1.0 catalogue (Table~\ref{tab:uvcounterparts}).\\
\indent Although the source is quite bright, it seems that only the inner 
5 pixels may be affected by pile up. We therefore extracted the spectrum 
from an annulus of 5 pixel inner radius and 30 pixel outer radius.
The spectrum is well-fitted with an absorbed power-law (\chisq=0.99 for 154 dof) with parameters 
that are well constrained. The value of \nh\ is lower than what 
is obtained through the LAB survey \citep{kalberla05}. The latter value is measured from 
a pointing that is $\sim$17\arcmin\  from the best position of the source, and spatial 
variations of the Galactic \nh\ could result in this apparent discrepancy. In a second 
run we, however, froze \nh\ to the best value returned by the LAB survey. We obtained
a good fit (\chisq=1.07 for 155 dof) and $\Gamma=2.04\pm0.04$ slightly softer 
than when leaving \nh\ free to vary. The parameters are clearly compatible with those
of  a Sey 1 AGN.

\subsection{\object{J12060+3818}}
This object was first reported in \citet{paltani08} \correc{who give a probability of only 59\% that 
the IGR source is real.} There is no X-ray source within the IBIS
error, but we notice a rather significant ($6.6\sigma$) source (\object{Swift J120617.2+381837}) 
at $\sim 6$\arcmin\ from the centre of the IBIS error.  \correc {First, one should note 
that \citet{paltani08} give an error of $\sim3.96$\arcmin, which is a bit underestimated 
compared to the expected 4.72\arcmin\  error of a $5.26\sigma$ source as obtained 
from \citet{gros03}. In this  case the \swift\ source is within the 
$97\%$ (i.e. $<3-\sigma$) of the IBIS confidence radius.  Now, given that only 1 $>3-\sigma$ source is 
detected over the whole detector,  the mean number of source  within a 6\arcmin\ radius 
from the IBIS position is $\sim0.26$. By assuming a Poisson distribution, we can estimate a probability of
19\% to find a  source by chance in this area.  Although chance association is non-negligible, } 
we study the possibility here that the sources are related. IGR J12060+3818
 has been associated to \object{2MASX J12055104+3819308} by \citet{paltani08}.
Since this IR source is not the IR counterpart to Swift J120617.2+381837, if the IGR and Swift sources 
are the same, then the former source is wrongly identified with the IR source in \citet{paltani08}.  
The error box of XRT contains two quasars, \object{SDSS J120617.35+381234.9}
at $z=0.8379\pm0.0011$ and \object{B2 1203+38} at z=0.8380. We consider it very likely  that those 
two objects are the same.  A possible optical and UV counterpart is detected in all filters 
(Table~\ref{tab:uvcounterparts}), indicating that the source is not absorbed.\\
\indent The X-ray spectrum is represented well by an absorbed power-law (C=36.6 for 42 dof). As 
the parameters are poorly constrained, and since the value of the absorption (and the detection 
of the source at UV wavelengths) is compatible with very little absorption in the direction of this source, 
we froze \nh\ to the value returned by the LAB survey in a second run. The photon index tends 
to a harder value ($\Gamma=1.8\pm0.4$), a value more compatible with those usually found in 
AGN. The extrapolated 20--60 keV flux is far below the value found by \citet{paltani08}, which
could indicate that the \swift\ and \integral\ sources are not related. If they are related, then 
this source could be significantly variable. \correc{Our results lead us to tentatively classify Swift J120617.2+381837 as an AGN 
most probably of type 1.  While we cannot completely rule out a chance association of the IGR and 
\swift\ sources, that AGN are known hard X-ray emitters and the 81\% probability of true association 
between the 2 sources also make us conclude that the sources are likely the same.}
  
\subsection{\object{J12070+2535}}
As for the previous object while there is no X-ray object in the error IBIS box  given by \citet{paltani08},
we note a possible underestimation of this positional uncertainty. According to \citet{gros03}, the IBIS error 
for a $5.37\sigma$ source is 4.64\arcmin. In that case, a $8.3\sigma$ source, \object{Swift J120705.3+253906}, 
is found within the error.  The XRT error box contains one 2MASS object, at a position compatible with a USNO B1.0 source, itself
detected at UV wavelengths with UVOT (Tables~\ref{tab:ircounterparts} and \ref{tab:uvcounterparts}). The XRT position is 
also compatible with two objects reported in NED, \object{SDSS J120705.29+253906.0} and 
\object{MAPS-NGP O\_377\_0077115}, which are very probably a unique object (separation of 1\arcsec).
Both these objects are extended and classified as galaxies. Our refined X-ray position rules out the association of the 
IGR source with \object{IRAS 12046+2554} (also identified as \object{LEDA 38453} in SIMBAD) suggested by \citet{paltani08}.\\
\indent The XRT spectrum of this source is well-fitted with an absorbed power-law (C=56 for 47dof). 
The spectral analysis shows some absorption in excess of the value of the line of sight; however, its 
level remains low, and this object is clearly not a member of the heavily obscured sources. The extrapolated 
20--60 keV flux is much lower than that reported in \citet{paltani08}, which, again, may indicate
that the sources are either not related or that we are observing a variable X-ray source. Here again
we cannot make more conclusions about the association of the two sources. We nevertheless
tentatively classify Swift J120705.3+253906 as an AGN, most probably one of type 1.

\subsection{\object{J13042$-$1020}}
\begin{figure*}
\centering
\epsfig{file=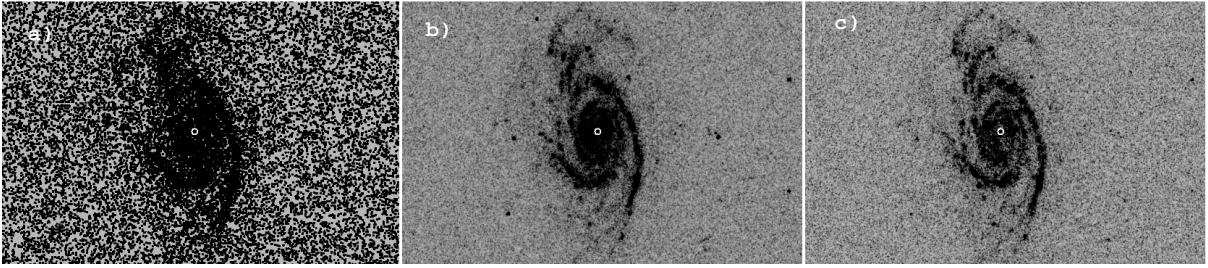,width=16cm}
\caption{9.6\arcmin$\times$6.0\arcmin  a): UVM2 b): UVW1 c): UVW2 UVOT images around IGR J13042$-$1020. 
The white circle shows the best XRT position for Src \#1, identifying it with the centre of the face-on spiral galaxy NGC 4939.}
\label{fig:13042}
\end{figure*}

The first mention of this object can be found in \citet{bird10}. There are four $>4\sigma$ XRT sources 
within the IBIS error box. Two of these (Src \#2 and Src \#4 in Table~\ref{tab:xray}) are coincident with 
HII regions related to the galaxy \object{NGC 4939} according to NED. We do not discuss them further here 
as it seems quite unlikely that they are at the origin of the hard X-ray source. Src \#3 has \object{2MASS J13040977$-$1019414}
in its error box. It is detected in the optical filters of UVOT but not at UV wavelengths. Src \#1, on the other hand, 
is positionally compatible with an extended 2 MASS source, reported as being NGC 4939 a z=0.010374 Sey 2. 
The X-ray position falls right on the centre of a face-on spiral Galaxy, as can be seen in Fig.~\ref{fig:13042}. We consider this object 
as the most promising counterpart. It is also reported in the 2nd XMM catalogue \citep{watson09} as 
\object{2XMM J130414.3$-$102021} at a position consistent with the \swift\ one. \\
\begin{figure*}
\centering
\epsfig{file=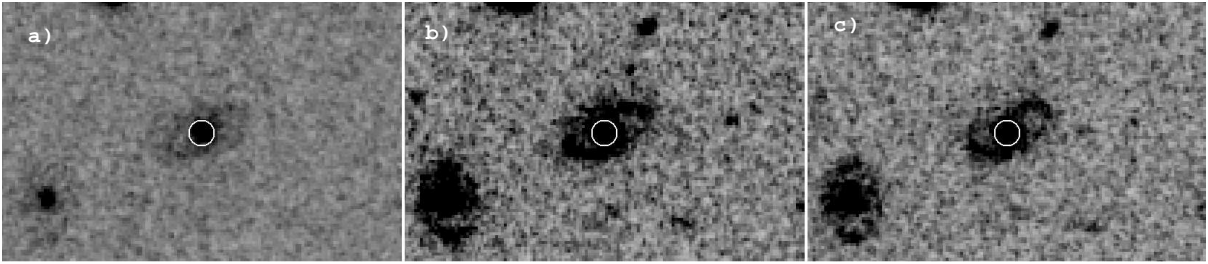,width=16cm}
\caption{2.3\arcmin$\times$1.5\arcmin a): UVM2 b): UVW1 c): UVW2 UVOT images around IGR J13412+3022. The white circle
shows the best XRT position, identifying it with the centre of the spiral galaxy Mrk 268.}
\label{fig:13412}
\end{figure*}
\indent To avoid contamination by Src \#2 and \#4, the region of extraction of the X-ray spectrum was slightly 
shifted.  The X-ray spectrum has a shape typical of an absorbed source showing a soft excess, \correc{and, indeed, a simple 
absorbed power-law does not provide an acceptable fit. The origin of soft 
excesses in AGN is subject to debate \citep[see e.g.][]{porquet04}. Amongst other possibilities it could, for example, be the signature 
of partial absorption of the inner flow, or the high-energy tail of thermal emission from the accretion disc. A spectral fit with 
a partial covering absorber and a power-law leads to a rather good fit (\chisq=0.95 for 9 dof). The value of the absorption 
 is clearly in excess of the value of the line of sight \nh\ (Table~\ref{tab:spectral}). The covering fraction is $97\pm1\%$, 
and the photon index is  typical of an AGN (Table \ref{tab:spectral}). The extrapolated 20--40 keV flux is 
about 4 times lower than the value obtained with IBIS \citep{bird10}. While this is not  definite proof against the 
partial covering model (could simply indicate that the source is variable), and since thermal emission is also a commonly 
suggested hypothesis for the origin of soft excesses \citep{pounds01,porquet04}, we also fitted the spectrum with 
an absorbed power-law and  a black body only absorbed by the intervening material on the line of sight (i.e.,  
\nh\ is frozen to $=0.033\times 10^{22}$~cm$^{-2}$). A  good fit is  also achieved (\chisq=0.79 for 
8 dof) with this model}. The parameters are poorly constrained, but seem to, 
however, indicate a rather hard spectrum ($\Gamma=0.1_{-0.8}^{+1.2}$) and some intrinsic absorption 
(\nh$=10_{-9}^{+7} \times 10^{22}$~cm$^{-2}$). 
The extrapolated 20--40 keV flux is $2.2\times10^{-11}$ \ergcms, which is  a factor of $\gtrsim6$ higher than  found with IBIS. 
The black body has a temperature 
kT$_{bb}=0.24_{-0.04}^{+0.05}$ keV and luminosity L$_{bb}=2.2\pm0.4 \times 10^{40}$ erg/s at z=0.0104. 
\correc{The black body temperature is compatible with those reported by \citet{porquet04}, and the luminosity is} 
compatible with an AGN nature for this source. \correc{In both cases (partial covering and 
black body), our results tend to further confirm the association of the XRT source with the IGR,} and 
the high absorption would argue in favour of a Sey 2.
 
\begin{table*}[htbp]
\centering
\caption{Results of the X-ray spectral analysis. Errors and upper limits are all given at 
the 90\% level.}\label{tab:spectral}
\begin{tabular}{lcllllll}
\hline\hline
Name & Net number & Galactic \nh $^\dagger$  & \nh\ & $\Gamma$& 2--10 keV flux$^\ddagger$ & \multicolumn{2}{c}{20--40 keV flux}\\
(IGR) &  of counts & $\times10^{22}$~cm$^{-2}$ & $\times10^{22}$~cm$^{-2}$  &   & \ergcms & \multicolumn{2}{c}{\ergcms}\\
         &                 &                                          &                                           &   &              & Extrapolated  & IBIS source \\
\hline
J02086$-$1742      & 931 & 0.01 & $<0.017$ & $1.55_{-0.08}^{+0.1}$ & $6.4\times10^{-12}$ & $6.3\times10^{-12}$ & $6.8\times10^{-12}$\\
J02524$-$0829$^\star$      & 617 & 0.04 & $12\pm3$ & $2.4\pm0.6$ & $1.2\times10^{-11}$ & $4.7\times10^{-12}$ & 3.1$\times10^{-11}$\\ 
J03184$-$0014      & 22   & 0.05 & $<1.1$ & $1.4_{-0.9}^{+1.5}$ & $1.5\times10^{-13}$ &$2.0\times10^{-13}$  & $2.9\times10^{-11}$ \\
J11457$-$1827$^\P$      & 4020 & 0.03 & $1.298_{-0.003}^{+0.01}$ & $1.91_{-0.04}^{+0.08}$ & $1.4\times10^{-11}$ &$1.1\times10^{-11}$ & $3.3\times10^{-11}$ \\
J12060+3818$^\P$       & 48    & 0.016 & $<0.13$ & $2.4_{-0.9}^{+1.2}$ & $1.5\times10^{-13}$ & $5.6\times10^{-14}$ & $3.1\times10^{-11}$\\
J12070+2535$^\P$         & 74    & 0.016 & $0.3\pm0.2$ & $1.9_{-0.5}^{+0.6}$ & $1.4\times10^{-12}$ &  $1.1\times10^{-12}$&  $1.4\times10^{-11}$\\
J13042$-$1020$^a$      & 268 & 0.033 & $29_{-6}^{+8}$& $2\pm0.2$& $4.0\times10^{-12}$& $1.7\times10^{-12}$ & $7.5\times10^{-12}$ \\
J13412+3022$^a$         & 90  & 0.014 & $30_{-8}^{+11}$ & $1.4_{-0.6}^{+0.2}$ & $8.0\times10^{-12}$& $9.5\times10^{-12}$& $9.1\times10^{-12}$\\
J14488$-$5942\#1  & 344 & 1.8    & $13_{-4}^{+6}$ & $2.8_{-0.9}^{+1.1}$ & $7.9\times10^{-12}$ & $7.1\times10^{-13}$ & $3.8\times10^{-12}$\\
\hspace*{0.55cm} " \#2 & 22 &"  & $<2.4$ & $2.2_{-1.6}^{+2.2}$ & $4.7\times10^{-14}$ &$7.9\times10^{-15}$ & "  \\
J23130+8608         & 20  & 0.052 & -- -- -- & $1.4_{-0.9}^{+0.7}$ & $7.0\times10^{-14}$ & $9.4\times10^{-14}$ & $1.4\times10^{-11}$\\
\hline
\hline
\end{tabular}
\begin{list}{}{}
\item[$^\dagger$]Value obtained from the LAB survey \citep{kalberla05}.
\item[$^\ddagger$]Flux corrected for the absorption.
\correc{\item[$^\star$]Hard fluxes are estimated over the 17--60 keV ranges.}
\correc{\item[$^\P$]Hard fluxes are estimated over the 20--60 keV ranges.}
\correc{\item[$^a$]Spectra fitted with a partial covering absorber and a power-law.}
\end{list}
\end{table*}

\subsection{\object{J13412+3022}}
The first mention of this object can be found in \citet{bird10}. The authors, however, give the source 
a wrong name (IGR J13415+3033) when considering the source coordinates obtained by IBIS. We therefore
propose to rename the source IGR J13412+3022.  The source is also referred to as \object{IGR J13415+3023}
in SIMBAD. There is one single bright XRT source within the IBIS error 
box. There are two objects in the XRT error box according to SIMBAD. One is \object{Mrk 268}, a Sey 2 galaxy
at z=$0.03986\pm0.00004$ according to NED, the other
 SN1994o an SN Ia. We disregard the second object as a possible counterpart to the IGR source and 
suggest that Mrk 268 is the true counterpart.  This object is reported in the IBIS 
catalogue of \citet{krivonos07}, but the position they obtained has a 6.5\arcmin\ offset compared to the XRT one. 
There are  two possible IR counterparts within the XRT error box. The first is the one reported in Table \ref{tab:ircounterparts}. It 
is an extended source and is clearly the IR counterpart to the Sey 2. The second one is \object{2MASS J13411114+3022410},  
a point source at 0.8\arcsec\ from the centre of the \swift\ position. We favour the first as the infrared counterpart to 
the high-energy source. The source is also clearly detected in the USNO B1.0 catalogue and 
in all UVOT filters (Table~\ref{tab:uvcounterparts}, Fig.~\ref{fig:13412}). Note the appearance of 
 spiral arm-like structures clear at shorter UV wavelengths, compatible with the classification of the 
host galaxy as a SBb as reported in NED.\\
\indent As for the previous IGR source, the spectrum, although of poor quality, may show two 
well separated broad structures. An absorbed power-law does not provide an acceptable fit, 
and large residuals are seen at very soft X-rays. \correc{Similar to the previous source, 
a fit with a partial covering absorber and a power-law provides an improvement over the simple 
absorbed power-law (C=81 for 45 dof). The source shows intrinsic absorption 
with a large covering fraction ($98_{-6}^{+2}\%$)  and a rather hard  
photon index (Table~\ref{tab:spectral}). With this model the extrapolated 20--40 keV flux is  \ergcms, a value
clearly compatible with the IBIS flux (Table~\ref{tab:spectral}).} Adding a black body  
\correc{(only absorbed by the intervening material on the line of sight) to a fully absorbed power-law also} leads
 to a good representation of 
the spectrum (C=71 for 44 dof). The black body has a temperature kT$_{bb}=0.31_{-0.09}^{+0.15}$ keV
and a luminosity L$_{bb}=2.9_{-1.0}^{+1.7}  \times 10^{41}$ erg/s at z=0.04, \correc{compatible with the values 
obtained for other AGN \citep[e.g.][]{porquet04}}. The power-law slope is, \correc{however}, very
poorly constrained \correc{and very soft ($\Gamma=4.3_{-2.4}^{+3.9}$) for an AGN}, but \correc{this model  also} 
indicates a high level of intrinsic absorption. In this case the extrapolated 20--40 keV 
flux is \correc{incompatible with} the IBIS flux reported in \citet{bird10}. However, freezing the photon index to a 
more common value of 1.8 leads to a 20--40 keV flux just slightly below what is reported in the 
4$^{th}$ IBIS catalogue. These results are clearly consistent with the source being a Sey 2 AGN.  
We therefore conclude that IGR J13412+3022 is the hard X-ray 
counterpart to Mrk 268, hence  a Sey 2.

\subsection{\object{J14488$-$5942}}
This source was first mentioned in \citet{bird10} as a transient. There are two X-ray sources 
within the IBIS error box. Src \#1 (\object{Swift J144843.3$-$594216}) has the highest significance.
Its X-ray error box contains a 2MASS but no USNO B1.0 source. It is not detected by the UVOT 
telescope with lower limits (based on the faintest object detected in the images) m$_{\rm{U}}>22.1$ and
m$_{\rm{UVW1}}>21.6$. This is expected given that the object is in the Galactic plane and tends 
to suggest that it lies at a rather large distance in our Galaxy. \\
\indent Src \#2  (\object{Swift J144900.5$-$594503}) has a rather low significance (Table~\ref{tab:xray}). It
has no counterpart in any of the online catalogues, although it is about 3\arcmin\ from \object{G 317.3-0.2-41.5}, a 
molecular cloud. An association of the two cannot be dismissed. It is not detected in the UVOT filters.  \\
 \indent  An absorbed 
power-law gives a marginally acceptable fit (\chisq=1.8 for 14 dof) to the spectrum of source \#1. This fit indicates 
that significant intrinsic absorption
shields the intrinsic emission of the source. An absorbed black body \correc{provides a good description of the }
spectrum as well with 
a temperature $kT=1.2_{-0.2}^{+0.3}$ keV and a  luminosity of $6\times10^{34}$ erg/s/($(\rm{d}/10\rm{kpc})^2$).
The extrapolated 20--40 keV flux \correc{obtained with the absorbed power-law model} is \correc{a factor of $\gtrsim5$} 
below the IBIS flux (Table~\ref{tab:spectral}). 
\correc{Flux difference would not be surprising if the \swift\ and IGR sources were associated, as 
the IGR is reported as a transient and is }therefore a  
variable source. In this case, the soft spectrum  and the black body fit 
may indicate that we are seeing a neutron star binary in quiescence.  \\
\indent\correc{The spectrum of source \#2 has a very low significance. An absorbed power-law provides 
a reasonable fit (C=24.5 for 20 dof), but the parameters are very poorly constrained (Table~\ref{tab:spectral}). 
An absorbed black body also provides a good fit (C=25.5 for 20 dof). In this case, we obtain
\nh$<1.8\times10^{22}$~cm$^{-2}$ (therefore at maximum consistent with the value of Galactic absorption 
on the line of sight), $kT=0.5_{-0.2}^{+0.5}$ keV and a luminosity $<6.3\times10^{33}$ erg/s/($(\rm{d}/10\rm{kpc})^2$).}\\
\indent \correc{\citet{bird10} classify this source as a transient, since it was only detected during Rev 520. The 
analysis of both possible X-ray counterparts indicates that they are likely to be Galactic sources (which is 
strengthened by their low galactic latitude),  and, in that case
most probably X-ray binaries. While it is not possible to further confirm  which (if
any) of the 2 X-ray sources is the true counterpart to the IGR, source \#1 is intrinsically absorbed, which is reminiscent 
of many of the IGR X-ray binaries.  We also note that the amplitude of the variations of the 20--40 keV flux between the extrapolated value 
obtained from our fits, and the maximum reported in \citet{bird10} is of $\sim 3\times10^4$ for source \#2. Again while this is 
not definite proofs that object \#2 is not the IGR, all those points, and the fact that source \#1 is the closest in
position to the IGR make us slightly  prefer source \#1 as the counterpart to the IGR source.  }
In either case, we conclude that IGR J14488$-$5942 has a Galactic origin \correc{and that, if it is indeed associated to 
source \#1},  could be a member of the absorbed X-ray binaries. 

\subsection{\object{J15283$-$4443}}
This source was discovered by \citet{2006ATel..865....1P}, and apart from its detection in a 
single \integral\ pointing, nothing is known about it. We detect a single faint X-ray source 
at a position compatible with that of IBIS. There is one IR, optical, and UV counterpart within 
the XRT error box (Tables~\ref{tab:ircounterparts} and\ref{tab:uvcounterparts}). The source is know 
as \object{TYC 7847-975-1} and classified as a star. It is very bright in the UVOT filters and saturates 
the V, B, and U filters. The magnitudes in these filters should then be taken with caution. \\
\indent The spectrum has too few counts (9 net cts) to be exploitable. \correc{However, to compare 
the possible flux from the \swift\ source in the \integral\ range, we  assumed a power-law 
spectrum whose photon index was frozen to different values, and let normalization free to vary. The highest 
extrapolated 17--40 keV flux is obtained with $\Gamma=0.2$. It is, however, still a factor of $\sim1000$
lower than the flux obtained with IBIS by \citet{2006ATel..865....1P}. }We therefore can only conclude 
something about the Galactic nature of this source if the IGR source is associated to TYC 7847-975-1. The TYC source has a 
high proper motion, which indicates a probable nearby object.  

\subsection{\object{J23130+8608}} 
This source was first reported in \citet{bird07}, but is absent in the last version of the 
IBIS catalogue. We found one faint X-ray source within the IBIS error box. The positional uncertainties
of the XRT and the 2MASS catalogue render  the IR and X-ray sources marginally compatible (Table~\ref{tab:xray} and 
\ref{tab:ircounterparts}). The IR position is also compatible with that obtained at UV wavelengths
with UVOT. The optical counterpart mentioned in Table \ref{tab:uvcounterparts} is more than 2\arcsec\ away from
the position of the IR (and UV) source, which indicates that these sources are probably not related.\\
\indent The X-ray spectrum has  very low statistical quality and contains no count above 3 keV. As the 
parameters are very poorly constrained, and given the probable detection of the source at UV wavelengths, 
it  probably has a very low intrinsic absorption. This point is further confirmed by the low value taken 
by \nh\ when fitted with an absorbed power-law. A simple power-law provides an adequate fit (C=7 for 14 dof).
A simple black body also provides a good fit (C=6.3 for 14 dof), with kT$=0.3_{-0.1}^{+0.2}$~keV and 
L$=4\pm2 \times10^{32}$ erg/s/($(10\rm{kpc})^2$). With the results in hand we cannot conclude anything more about
the nature of the source, although the high Galactic latitude would tend to dismiss a Galactic compact object 
such as an X-ray binary. We do not exclude the \swift\ and \integral\ sources being unrelated. 

\section{Summary and conclusions}
In this paper, we reported the \swift\ X-ray analysis of the field of thirteen IGRs that  still lacked 
an arcsec accurate position. The refined X-ray positions provided by the \swift\ observations 
(Table~\ref{tab:xray}) allowed us to pinpoint the possible IR, optical, and UV counterparts 
in most of the cases. We also analysed the X-ray spectra of the sources and used these results 
as additional arguments to confirm or refute the association of the \swift\ source with the \integral\
one. This also helped us to tentatively give a possible classification for the X-ray source.
Table~\ref{tab:results} reports the conclusions of our analysis 

\begin{table}[htbp]
\caption{Summary of the possible type for each counterpart of the thirteen sources, obtained 
through our analysis.}
\label{tab:results}
\begin{tabular}{ll} 
\hline\hline             
Name &   Type \& Comment\\
(IGR) &     \\ 
\hline
J02086$-$1742 & AGN, possibly Sey 1 \\
J02524$-$0829 & Sey 2 AGN at z=0.016721 \\
J03184$-$0014 & possible AGN, \swift\ and IGR associated?\\
J08023$-$6954 & RS CVn \\
J11427+0854    & ?, possible spurious IGR\\
J11457$-$1827 & Sey 1 AGN at z=0.0329 \\
J12060+3818    & QSO, possibly Sey 1, \swift\ and IGR associated?\\
J12070+2535    & AGN, possibly Sey 1,  \swift\ and IGR associated?\\
J13042$-$1020\#1 & AGN, possibly Sey 2 \\
J13412+3022        & Sey 2 AGN at z=0.03986  \\
J14488$-$5942\#1 & probable XRB\\
J15283$-$4443      & Galactic source, \correc{\swift\ and IGR associated?}\\
J23130+8608         & ? , \swift\ and IGR associated?\\
\hline
\hline
\end{tabular}
\end{table}

We can summarise our results as follows.\\
\begin{itemize}
\item We identify IGR J02086$-$1742, IGR J12060+3818, IGR J12070+2535,
IGR J13042$-$1020, and IGR J13412+3022 as AGN. We, however, do question 
the associations of IGR J12060+3818 and IGR J12070+2535 with the X-ray counteparts 
we found. We suggest that IGR J02086$-$1742 is a possible Sey 1 and that 
IGR J13042$-$1020 is a possible Sey 2. Our analysis permits us to clearly 
identify IGR J13412+3022 as a Sey 2. 
\item We confirm the previously proposed associations of 
IGR J02524$-$0829, IGR J08023$-$6954, and IGR J11457$-$1827. 
These objects are respectively classified as a Sey 2 AGN, 
an RS CVn star, and a Sey 1 AGN.
\item We classify IGR J14488$-$5942 as a probable XRB,
\item Apart from classifying  IGR J15283$-$4443 as a Galacric source, we cannot 
 conclude much more about the nature of this source .
\item We provide new data for IGR J03184$-$0014. We confirm the presence of a source 
found in Paper 2, but we discuss its possible association with the IGR source. We provide
a new SDSS identification for the counterpart, which is classified as a galaxy. The X-ray 
source is therefore an AGN, which leads us to tentatively associate it with the IBIS source, 
although we notice  that the latter could be spurious. 
\item We are not able to give a classification for IGR J11427+0854 and IGR J23130+8608.
In both cases, the association of the X-ray and the IGR can be questioned. We further 
question the genuineness of the former IGR source.
\end{itemize}
Caution is needed with these proposed identifications, as definitive conclusions
will only come from optical/IR spectroscopy. We  are, however, confident wbout the objects
proposed as AGN, because they come from  identifying  extended counterparts
within the XRT error box. Note that IGR J13412+3022 is also a known  Sey 2 object. 

\begin{acknowledgements}
JR thanks S. Soldi,  I. Caballero, and P. Ferrando for useful discussions.  
JAT acknowledges partial support from a NASA INTEGRAL Guest Observer INTEGRAL grant NNX08AX91G. 
AB acknowledges support from a NASA Chandra grant GO8-9055X. 
We warmly thank the referee for fruitful comments that helped to improve this 
paper. 
We acknowledge the use of data collected with the \swift\ observatory.
This research has made use of the USNOFS Image and Catalogue Archive
operated by the United States Naval Observatory, Flagstaff Station
(http://www.nofs.navy.mil/data/fchpix/)
 This research  made use of the SIMBAD database, operated at the CDS, Strasbourg, France.
It also made use of data products from the Two Micron All Sky Survey, which 
is a joint project of the University of Massachusetts and the Infrared Processing 
and Analysis Center/California Institute of Technology, funded by the National 
Aeronautics and Space Administration and the National Science Foundation.
This research  made use of the NASA/IPAC Extragalactic Database (NED), which is 
operated by the Jet Propulsion Laboratory, California Institute of Technology, under 
contract with the National Aeronautics and Space Administration. 

\end{acknowledgements}
\bibliographystyle{aa}

%\bibliography{../../BiblioFull}

\end{document}